\documentclass[pre,aps,twocolumn]{revtex4}

\usepackage{graphicx}
\usepackage{amsmath,empheq}
\usepackage{amssymb}
\usepackage{ifthen}
\usepackage{booktabs}
\usepackage{color}

\usepackage{graphicx}% Include figure files
\usepackage{dcolumn}% Align table columns on decimal point
\usepackage{bm}% bold math
\usepackage{longtable}

\newcommand{\bb}{\begin{equation}}
\newcommand{\ee}{\end{equation}}
\newcommand{\ba}{\begin{eqnarray*}}
\newcommand{\ea}{\end{eqnarray*}}
\newcommand{\rhor}{\rho({\bf r})}
\newcommand{\dd}{{\rm d}}
\newcommand{\rr}{{\mathbf r}}
\newcommand{\dr}{{\rm d}{\bf r}}

\begin{document}

\title{Asymptotic properties of bridging transitions in sinusoidally-shaped slits}

\author{Alexandr \surname{Malijevsk\'y}}
\affiliation{{The Czech Academy of Sciences, Institute of Chemical Process Fundamentals,  Department of Molecular Modelling, 165 02 Prague, Czech
Republic;}{Department of Physical Chemistry, University of Chemical Technology Prague, 166 28 Prague, Czech Republic}}
 \author{Martin \surname{Posp\'\i\v sil}}
\affiliation{{The Czech Academy of Sciences, Institute of Chemical Process Fundamentals,  Department of Molecular Modelling, 165 02 Prague, Czech
Republic;}{Department of Physical Chemistry, University of Chemical Technology Prague, 166 28 Prague, Czech Republic}}

\begin{abstract}

\noindent We study bridging transitions that emerge between two sinusoidally-shaped walls of amplitude $A$, wavenumber $k$, and mean separation $L$.
The focus is on weakly corrugated walls to examine the properties of bridging transitions in the limit when the walls become flat. The reduction of
walls roughness can be achieved in two ways which we show differ qualitatively: a) By decreasing $k$, (i.e., by increasing the system wavelength),
which induces a continuous phenomenon associated with the growth of  bridging films concentrated near the system necks, the thickness of with the
thickness of these films diverging as $\sim k^{-2/3}$ in the limit of $k\to0$. Simultaneously, the location of the transition approaches that of
capillary condensation in an infinite planar slit of an appropriate width as $\sim k^{2/3}$; b) in contrast, the limit of vanishing walls roughness
by reducing $A$ cannot be considered in this context, as there exists a minimal value $A_{\rm min}(k,L)$ of the amplitude below which bridging
transition does not occur. On the other hand, for amplitudes $A>A_{\rm min}(k,L)$, the bridging transition always precedes global condensation in the
system. These predictions, including the scaling property $A_{\rm min}\propto kL^2$, are verified numerically using density functional theory.
\end{abstract}

\maketitle

\section{Introduction}

Capillary condensation is a well-known phenomenon referring to the condensation of a gas due to the presence of confining walls. Perhaps the simplest
example of this is condensation inside a narrow slit formed by a pair of parallel planar walls that are a distance $L$ apart. If the walls are
``hydrophilic'', meaning  they preferentially adsorb liquid,  the confined fluid may condense under  thermodynamic conditions, for which the stable
phase of the bulk fluid is still a gas. This shift in the phase boundary  can be expressed in terms of the difference $\delta\mu_{cc}(L)=\mu_{\rm
sat}-\mu_{cc}(L)$ between the chemical potentials corresponding to capillary condensation, $\mu_{cc}(L)$, and bulk condensation, which occurs along
the saturation line $\mu_{\rm sat}(T)$, where $T$ is the subcritical temperature.

Macroscopically, the shift can be described by the classical Kelvin equation
  \bb
 \delta\mu_{cc}(L)=\frac{2\gamma\cos\theta}{L\Delta\rho}\,,  \label{kelvin_slit}
  \ee
where $\gamma$ is the liquid-gas surface tension, $\theta$ is Young's contact angle and $\Delta\rho=\rho_l-\rho_g$ is the difference in the one-body
densities of the bulk liquid and bulk gas. According to the Kelvin equation, capillary condensation occurs when $\theta<\pi/2$, in which case
$\delta\mu_{cc}(L)>0$; this corresponds to the case of ``hydrophilic'' walls recalling Young's law
 \bb
  \gamma_{wg}=\gamma_{wl}+\gamma\cos\theta\,,  \label{young}
 \ee
where $\gamma_{wg}$ and $\gamma_{wl}$ are the surface tensions of the wall-gas and the wall-liquid interfaces, respectively.

Conversely, if the walls are ``hydrophobic'', meaning $\theta>\pi/2$, then $\delta\mu_{cc}(L)<0$ and the system exhibits capillary evaporation at the
chemical potential above $\mu_{\rm sat}(T)$, which is essentially the same phenomenon as capillary condensation with the roles of liquid and gas
being swapped.

A significant step forward in understanding capillary condensation and related phenomena was made in the 1980's, when Evans et al. placed the subject
within the context of the statistical mechanics of inhomogeneous fluids \cite{evans90, hend}. This approach not only provided a more fundamental
foundation for Kelvin's equation but also paved the way for its extension to a more microscopic level by considering the nature of molecular
interactions. Formulated within the grand canonical ensemble, the approach is based on treating  a low-density (gas-like) state and a high-density
(liquid-like) state separately and comparing their stability. More specifically, for the model of an infinitely long slit, the grand potential per
unit area is expressed as a sum of the volume and surface contributions, assuming that the surface properties of the walls can be described as if the
walls were isolated. Thus, the grand potential per unit area of the low-density state can thus be written as:
 \bb
 \Omega_g=-pL+2\gamma_{wg}\,,  \label{omegag}
 \ee
where $p$ is the pressure of the bulk reservoir, assumed to be in the gas state, i.e., $p<p_{\rm sat}$, where $p_{\rm sat}$ is the vapor pressure of
the bulk fluid at the given temperature. Similarly, for the liquid state, we can write
 \bb
 \Omega_l=-p^\dagger L+2\gamma_{wl}\,,  \label{omegal}
 \ee
where $p^\dagger$ is the pressure of the metastable bulk liquid that corresponds to the given values of $\mu$ and $T$ which, together with the system
size $L$, define the appropriate grand canonical ensemble.

The Kelvin equation (1) is obtained by balancing $\Omega_l$ and $\Omega_g$  using  Young's equation (\ref{young}) and the expansion
$p-p^\dagger\approx|\mu-\mu_{\rm sat}|\Delta\rho$, which is accurate not too far from saturation. However, this approach allows for substantial
extensions beyond this result. For instance, in addition to predicting the shift of the phase boundary, it can determine the shift in the critical
temperature $T_c$ due to the confinement, which the Kelvin equation itself does not account for. This can be achieved by invoking scaling arguments
and considerations of bulk criticality  \cite{NF81, NF83}.

Moreover, the Kelvin equation can be modified on the mesoscopic scale by accounting for the impact of thick liquid-like layers adsorbed at the walls
near saturation, particularly when the walls are completely wet, $\theta=0$ by making a link with the theory of wetting phenomena. This involves the
effective reduction of the slit width, which depends on the range of microscopic forces, known as Derjaguin's correction \cite{derj}. Importantly,
microscopic models for wetting phenomena \cite{sul79, sul81}, i.e. when only one wall is involved, have been applied to gain a microscopic insight
into the generic properties of phase equilibria of confined fluids \cite{evans86}. Numerous microscopic tests, including computer simulations
\cite{binder08} and classical density functional theory (DFT) studies \cite{evans90, gelb}, as well as recent  experimental results \cite{geim},
confirm the surprisingly high quantitative accuracy of these predictions even at the atomistic scale. Note that the similar arguments can also be
applied to formulate Kelvin's equation for cylindrically shaped walls, although in this case capillary condensation is inevitably rounded due to the
quasi-one-dimensional character of the system \cite{privman}.

The solid theoretical foundations for the equilibrium theory of capillary condensation in simple fluids (and their mixtures) within the simplest
model pores provide a basis for investigating more complex systems, which present new theoretical challenges and technological opportunities. These
systems include, for instance, specific equilibrium and transport phenomena that emerge in extremely narrow pores, sometimes referred to as
single-digit nanopores, where excluded volume effects have a significant impact \cite{faucher, aluru}. Examining the role of confinement in the
self-organisation of soft matter systems is another rapidly evolving field of research \cite{araujo}. Another direction extending the early studies
focuses on more complex geometries of the confining walls, which exert potentials that vary at least in two dimensions. In these cases, even the
equilibrium phase behaviour can be substantially more complex  due to additional interfacial and finite-size effects that are absent in simpler
planar or cylindrical geometries. Beside their fundamental significance, understanding the equilibrium behaviour of such systems is also  an
important prerequisite for studying their non-equilibrium properties, which are highly relevant to modern nanotechnologies.

In this paper, we focus on a macroscopic description of bridging transitions occurring in narrow slits formed by weakly undulated walls that are
arranged in a reflection symmetric manner. Due to the periodically varying width of the slit, the system may, under certain conditions, allow for the
formation of a linear array of bridges connecting opposite sections of the walls -- i.e.,  local condensation at the constricted regions of the slit
prior to global condensation. Note that global condensation cannot occur continuously (except in cases where the maximal separation of the walls is
smaller than the bulk correlation length) but must proceed via a first-order transition, as the fusion of bridging films is associated with a change
in (surface) free energy. Generally, bridging is an significant phenomenon,  frequently observed in soft matter systems, where it can act as an
effective solvent-mediated force between  pairs of colloids, thereby stabilizing various microstructures \cite{israel} and is closely related to
phenomena such as coagulation, flocculation, sintering, and others. The conditions under which capillary bridging films are formed (or break) and
their properties have been  extensively studied both experimentally \cite{mason, beysens, willett, kohonen, fournier, gogelein} and theoretically
\cite{macro_rings, macro_vogel, macro_willet, yeomans, yeomans2, bauer, andrienko, grof, hopkins, dutka, cheng, mal, mal_par, farmer, labbe, vasil,
vasilyev, mal24}.

The objective of the present work is to address some fundamental questions related to bridging in sinusoidally-shaped slits and particularly to
investigate its properties  when the corrugation of the confining walls is gradually reduced. While it is evident that bridging must ultimately turn
to capillary condensation in a planar slit, we aim to describe this process and elucidate the connection between these two types of phase
transitions. Since the corrugation of the walls can be reduced in two distinct ways -- either  by decreasing their amplitude or by increasing their
wavelength -- we examine both routes separately and compare them to uncover any potential differences.

To achieve this, we proceed as follows: In Section II, we recall the main properties of capillary condensation between a pair of planar, parallel
plates of \emph{finite} length. In Section III, we demonstrate how this simple model can be utilized to describe bridging transitions between
corrugated walls. This general framework is then specialized in Section IV by considering sinusoidally-shaped walls.  Here, we formulate the full
Kelvin-like equation, along with its linearized version suitable for weakly corrugated walls, which is subsequently a subject of an asymptotic
analysis. The  predictions obtained are then tested against a microscopic density functional theory, the model of which is formulated in Section V.
The comparison of results is presented in Section VI. Finally, the paper concludes with Section VII, where we summarize the findings and discuss
possible extensions of the present work.

%\begin{figure}[htb]
%  \includegraphics[width=\columnwidth]{}
%  \caption{} \label{x-alpha_inf}
%\end{figure}

\section{Capillary condensation in finite slits}

Prior to considering bridging transitions, we begin by briefly recalling the main aspects of capillary condensation  in open slits. These slits are
formed by two parallel plates of finite length $H$ and are separated by a distance denoted as $D$. Later, we will demonstrate that this model is
highly relevant for  describing bridging between two surfaces of arbitrary geometry. The walls are assumed to be of macroscopic depth, so that the
system is invariant with respect to translation along the walls. Specifically, let us consider the case where the contact angle of the walls is
$\theta<\pi/2$, meaning that the walls preferentially adsorb the liquid phase.  Thus, we expect that the fluid inside such a slit condenses at a
chemical potential $\mu_{cc}(D,H)<\mu_{\rm sat}$, i.e., before the bulk gas phase does.

On a macroscopic level, the grand potential per unit length for a low-density, gas-like phase can be expressed as:
 \bb
 \Omega_g=-pDH+2\gamma_{wg}H\,,
 \ee
where $p(\mu,T)$ is the pressure of the bulk gas. In the high-density, liquid-like state, two symmetric menisci of arc-length $\ell$ form to separate
the condensed phase from the ambient gas. The macroscopic grand potential per unit length in this state is given by:
  \bb
 \Omega_l=-p^\dagger (DH-2S)+2\gamma_{wl}H+2\gamma\ell\,,
  \ee
where $p^\dagger(\mu,T)$ is the pressure of the bulk (metastable) liquid and $S$ is the area between each meniscus and the open end of the slit. The
menisci have a radius of curvature $R=\gamma/\delta p$, where $\delta p=p-p^\dagger$, and must connect the walls at their ends with the \emph{edge
contact angle} $\theta_e$, rather than the equilibrium Young contact angle $\theta$ (see Fig.~\ref{fig_map}b).

The edge contact angle is connected with the system geometry via the expression $D/2=R\cos\theta_e$, which allows to express the location of the
transition, $\delta\mu_{cc}(D,H)=\mu_{\rm sat}-\mu_{cc}(D,H)$,  analogously to Eq.~(\ref{kelvin_slit})
 \bb
  \delta\mu_{cc}(D,H)=\frac{2\gamma\cos\theta_e}{\Delta\rho D}\,, \label{kelvin_fs}
 \ee
where the only difference is that Young's contact angle $\theta$ is replaced by the edge contact angle $\theta_e$. The edge contact angle is itself a
function of the pressure (or chemical potential), and at the point of capillary condensation, where $\Omega_g=\Omega_l$,  it is implicitly determined
by the relation \cite{our_slit}
 \bb
  \cos\theta_e=\cos\theta-\frac{D}{2H}\left[\sin\theta_e-\sec\theta_e\left(\frac{\pi}{2}-\theta_e\right)  \right]\,. \label{thetae_fs}
 \ee

Eq.~(\ref{thetae_fs}) is only valid for slits with $H>H_{\rm min}$ where $H_{\rm min}=D\sec\theta$ is the minimal wall height still that still allows
for capillary condensation. For $H<H_{\rm min}$, we have $\delta\mu_{cc}(D,H)<0$, indicating that capillary condensation is preceded by bulk
condensation. Similarly, capillary evaporation is preceded by bulk evaporation if $\theta>\pi/2$, in which case $\delta\mu_{cc}(D,H)>0$ for $H<H_{\rm
min}$.

 Eq.~(\ref{thetae_fs}) implies that for finite slits $\theta_e>\theta$, with the difference asymptotically vanishing as
 \bb
\theta_e-\theta\approx \left\{\begin{array}{ll} \left(1+\frac{\pi-\theta}{\sin\theta}\right)\frac{ D}{2H}& \theta>0\,,\\
 \sqrt{\frac{\pi}{2}\frac{D}{H}}& \theta=0\,, \end{array}\right.  \label{thetae_as}
 \ee
for $H\gg D$. This reveals a nonanalytic asymptotic behaviour of $\theta_e$ in the case of completely wet walls ($\theta=0$). It follows that
condensation in finite slits occurs closer to saturation than in infinitely long slits. The difference in the chemical potential,  valid for a high
aspect ratio \cite{open_slits}, is given by:
  \bb
  \delta\mu_{cc}(D)-\delta\mu_{cc}(D,H)\approx\frac{\alpha_1\gamma}{H}\,,\;\;\;(H\gg D)
  \ee
  where $\alpha_1=(\pi/2-\theta)\sec\theta+\sin\theta$.

\section{Macroscopic conditions for bridging transition} \label{macro}

In this section, we formulate the general conditions for bridging transitions between non-planar surfaces. We assume that the surfaces are smooth,
such that their local height (relative to the horizontal) can be expressed in  Cartesian coordinates as $\pm z_w(x)$, where the differentiable
function $z_w(x)$ is assumed to be even, with a global minimum located at $x=0$. As in the previous models, we  consider surfaces that are
macroscopically deep, implying that the system is translation-invariant along the remaining $y$ coordinate.

\begin{figure*}[htb]
\includegraphics[width=15cm]{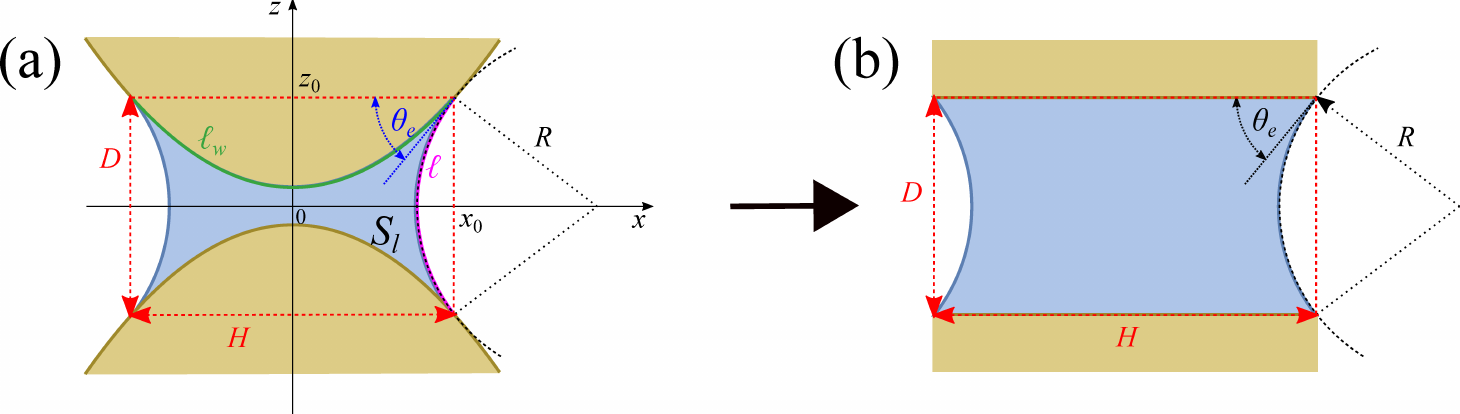}
\caption{ a) Scheme of a bridge formed between two nonplanar, completely wet walls ($\theta=0$). The preferentially adsorbed phase, shown in blue, is
separated by two menisci with  Laplace radius $R$ that tangentially connect the walls at coordinates $[\pm x_0,\pm z_0]$. The walls are assumed to be
macroscopically deep, resulting  in translational symmetry along the $y$-axis (perpendicular to the $x$-$z$ plane shown). The bridging film has a
surface area $S_l$, while  $\ell$ denotes the arc length of each meniscus,   and $\ell_w$  represents the contact length of the film with the walls.
b) The system can be mapped to a planar slit geometry with a finite length $H=2x_0$ and width $D=2z_0$. In this effectively equivalent model, the
menisci form an edge contact angle $\theta_e$, which deviates from $\theta$ (zero in this case) due to the wall curvature, as described by
Eq.~(\ref{thetae_theta}).} \label{fig_map}
\end{figure*}

The bridging transition occurs when grand potentials (per unit length and per period) of the gas-like and bridged-like phases balance, which can be
expressed as
 \bb
 \delta pS_l+2\gamma\ell-2\gamma\cos\theta\ell_w=0\,. \label{omegaex}
 \ee
 Here, the first term  corresponds to the free-energy cost due to the formation of the bridge phase, which is metastable in the bulk; the second
term represents the free-energy cost associated with the formation of two menisci of arc-length $\ell$; and the last contribution accounts for  the
interfacial free energy due to the contact of the walls (of the total length $2\ell_w$) with the bridge, where Young's law has been applied (cf.
Fig.~1a).

It is convenient to introduce, by analogy with the description of capillary condensation in finite slits, the angle $\theta_e$ subtended by the
tangent to the meniscus at its contact point with the wall and the horizontal line (see Fig.~\ref{fig_map}a). The analogy can be made even closer by
defining $H=2x_0$ and $D=2z_0$, i.e. the length and width of the rectangle surrounding  the bridge, thereby mapping our system to that of a finite
slit, where $\theta_e$ plays the role of the edge contact angle as defined in the previous section (see Fig.~\ref{fig_map}b). In this manner, the
location of the bridging transition can be expressed in a way formally identical to Kelvin's equation (\ref{kelvin_fs}) for a finite slit
 \bb
  \delta\mu_b=\frac{2\gamma\cos\theta_e}{\Delta\rho D}\,, \label{kelvin_bridge}
 \ee
indicating the departure in the chemical potential at the bridging transition from bulk coexistence. Here, $\delta\mu_b$ is a functional of $z_w(x)$
through the geometric measures characterizing the bridge shape:
 \bb
 \ell_w=2\int_0^{x_0}\sqrt{1+[z_w'(x)]^2}dx\,,
 \ee

\bb
 S_l=S-(\pi-2\theta_e)^2R^2+RD\sin\theta\,,
 \ee
 where
 \bb
  S=4\int_0^{x_0} z_w(x) dx\,,
 \ee

 and
 \bb
  \ell=\left(\frac{\pi}{2}-\theta_e\right)R\,.
 \ee
%where $S=4\int_0^{x_0} z_w(x) dx$ is the total available volume between the surfaces in the pertinent region.
% recalling that
%$R=\gamma/(\Delta\rho\delta\mu)$ is the Laplace radius of curvature of the menisci at the chemical potential $\mu=\mu_{\rm sat}-\delta\mu$.
%Now, in contrast to $\delta\mu_{cc}^(D,H)$ requiring the evaluation of $\theta_e$ from Eq.~(\ref{thetae_fs}) for the fixed parameters $H$ and $D$,
%the present task to obtain $\delta\mu_b$ requires the knowledge of the meniscus location $x_0$ from the condition $\Omega^{\rm ex}=0$ with
% \bb
%  \theta_e=\tan^{-1}(z_w'(x_0))+\theta\,.
% \ee
 To highlight the analogy with open slits,  Eq.~(\ref{omegaex})  can be rewritten as follows:
 \bb
  \bar{r}\cos\theta_e=r\cos\theta-\frac{D}{2H}\left[\sin\theta_e-\sec\theta_e\left(\frac{\pi}{2}-\theta_e\right)  \right]\,, \label{thetae_bridge}
 \ee
 where the angle $\theta_e$ is related to the shape of the walls by
  \bb
  \theta_e=\tan^{-1}(z_w'(x_0))+\theta\,. \label{thetae_theta}
 \ee
 Eq.~(\ref{thetae_bridge})  differs from Eq.~(\ref{thetae_fs}) only by the presence of the roughness parameters:
  \bb
  r=\frac{\ell_w}{H}
  \ee
  and its two-dimensional analog
  \bb
  \bar{r}=\frac{S}{HD}\,,
  \ee
  both of which reduce to unity for flat geometries.

\section{Bridging transition in sinusoidal slits}

We now apply the general results from the previous section to describe  bridging transitions occurring between sinusoidally undulated walls. The
local height of the walls relative to the horizontal $x$-axis is given by $\pm z_w(x)$, with
 \bb
 z_w=\frac{L}{2}-A\cos(kx)\,, \label{sin}
 \ee
where $L$ is the mean separation between the walls, $A$ is their amplitude, $\lambda$ is their wavelength and $k=2\pi/\lambda$ is the wavenumber. The
location of the bridging transition in such a geometry, fully characterized by the horizontal extent of the bridge $x_0$ at the point of transition,
is determined by the condition $\Omega^{\rm ex}=0$, obtained from Eq.~(\ref{omegaex}) (or equivalently, (\ref{thetae_bridge}))
 \begin{widetext}
 \begin{equation}
\left[2Lx_0-\frac{4A}{k}\sin(kx_0)\right]\cos\theta_e=2Dx_0+\frac{A^2k^2D}{4}\left[x_0-\frac{\sin(2kx_0)}{2k}\right]
-\frac{D^2}{2}\left[\sin\theta_e+\left(\frac{\pi}{2}-\theta_e\right)\sec\theta_e\right]\,, \label{kelvin_sin_full}
 \end{equation}
 \end{widetext}
where $D=L-2A\cos(kx_0)$.

Focusing on the case of completely wet walls ($\theta=0$), we can substitute for $\cos\theta_e=1/\sqrt{1+A^2k^2\sin^2(kx_0)}$,
$\sin\theta_e=Ak\sin(kx_0)/\sqrt{1+A^2k^2\sin^2(kx_0)}$, and $\theta_e=\tan^{-1}[Ak\sin(kx_0)]$, to obtain a nonlinear equation for $x_0$, which can
 be solved numerically. However, an analytic approach is feasible for weakly corrugated walls, such that $A^2k^2\ll1$. In which limit,
Eq.~(\ref{kelvin_sin_full}) simplifies to
 \bb
 4A\sin(kx_0)=4Akx_0\cos(kx_0)+\pi k\left[\frac{L}{2}-A\cos(kx_0)\right]^2\,. \label{kelvin_sin}
 \ee

It is convenient to introduce the dimensionless variables $\alpha=A/L$, $\beta=\pi kL/4$ and $u=kx_0$, in terms of which Eq.~(\ref{kelvin_sin})
becomes
 \bb
 \alpha\sin u=\alpha u\cos u+\beta\left(\frac{1}{2}-\alpha\cos u\right)^2\,, \label{kelvin_sin2}
 \ee
 which is subject to the condition $\alpha^2\beta^2\ll1$. We will examine this equation under three different regimes:

\begin{enumerate}

\item ${\mathbf \beta\gg\alpha}$

In this regime, Eq.~(\ref{kelvin_sin2}) is dominated by the last term unless its prefactor vanishes. The balance between the left and right sides of
the equation can only occur if
 \bb
 \alpha\cos u=\frac{1}{2}\,,
 \ee
which implies that $\alpha\ge 1/2$ violating the assumptions $\beta\gg\alpha$ and  $\alpha^2\beta^2\ll1$. Thus, bridging in sinusoidal slits requires
that the amplitude  $A$ is at least of the same order as $k$, reducing to one of the two remaining cases.

\item ${\mathbf \beta\sim\alpha}$

When $\alpha$ and $\beta$ are comparable, all terms in Eq.~(\ref{kelvin_sin2}) are relevant but the equation can be simplified as follows:
 \bb
 \sin u=u\cos u+\frac{\gamma}{4}\,, \label{kelvin_sin3}
 \ee
where $\gamma=\beta/\alpha$ and where the higher-order terms were neglected. The maximal value of $\gamma$ for which Eq.~(\ref{kelvin_sin3}) has a
solution occurs near $u\approx\pi/2$, beyond which bridges become unstable.  Thus, the maximal $\gamma$ is $\gamma_{\rm max}=4$, indicating that the
\emph{minimal} amplitude $A_{\rm min}$ required for bridging  satisfies
 \bb
 A_{\rm min}=\frac{\pi}{16}k L^2\,. \label{A_min}
 \ee
This confirms that $A_{\rm min}$ is indeed of the same order as $k$.

\item ${\mathbf \beta\ll\alpha}$

Finally, we consider asymptotic limit $k\to0$ with fixed $A$.  From Eq.~(\ref{kelvin_sin2}) it follows immediately that $x_0$ must diverge in this
limit, and that the last term of the equation becomes a correction to the ``unperturbed'' equation $\sin u=u\cos u$, which has a solution $u=0$. This
implies that $x_0\sim k^{-\nu}$ as $k\to0$ with $0<\nu<1$. Expanding Eq.~(\ref{kelvin_sin2}) in $u$ up to  third order yields
 \bb
 u=ck^{\frac{1}{3}}+{\cal{O}}(k)\,, \label{asym_u}
 \ee
 where
 \bb
c=\left[\frac{3\pi}{4}\frac{\left(\frac{L}{2}-A\right)^2}{A}\right]^{\frac{1}{3}}\,.
 \ee
 Therefore,
 \bb
  x_0\sim  k^{-\frac{2}{3}}\,,\;\;\;({\rm as}\;k\to0)\,, \label{asym_x0}
 \ee
with the subdominant term of order unity.

\end{enumerate}

To summarize the present results, there exists a critical phenomenon associated with an increase in the wavelength of the walls  (while keeping their
amplitude fixed), which manifests as the divergence of the bridging film width $x_0$ according to Eq.~(\ref{asym_x0}). Alternatively, the roughness
of the walls can also be reduced by decreasing their amplitude. However, in this case,  there exists a certain limit, $A_{\rm min}(k,L)\propto kL^2$,
below which the bridging transition becomes unstable.

Furthermore, it is important to examine whether the bridging transition is metastable with respect to (and thus preceded by) capillary condensation,
i.e., the complete condensation of fluid in the entire volume of the system. For the bridging transition to be stable, it is necessary that
 \bb
 \frac{\cos\theta_e}{D}>\frac{1}{L}\,,
 \ee
 which, after substituting for $\theta_e$ and $D$, leads to the condition
 \bb
 kx_0<\frac{\pi}{2}\,.
 \ee

This immediately implies that for weakly corrugated walls the bridging transition is always stable, since it occurs at the chemical potential
(pressure) pertinent to a smaller Laplace radius than that  for capillary condensation (which is $R_{cc}=L/2+{\cal{O}}(A^2k^2)$).

Finally, it is straightforward to show, by substituting from Eq.~(\ref{asym_u}) into the Kelvin equation (\ref{kelvin_bridge}),  that in the limit as
$k\to0$, we obtain
 \bb
 \delta\mu_b=\delta\mu_{cc}(L-2A)\left(1-\frac{Ac^2}{L-2A}k^{\frac{2}{3}}\right)\,, \label{asym_dmu}
 \ee
which relates the location of the bridging transition to that of capillary condensation in a planar infinite slit of width $L-2A$. This result
suggests that bridging in a weakly sinusoidally corrugated slit occurs under the same conditions as in an infinite slit of effective width
$\tilde{L}$,
 \bb
 \delta\mu_b=\delta\mu_{\rm cc}(\tilde{L})\,,\label{covariance}
 \ee
 where
 \bb
  \tilde{L}=L-2A+ac^2k^{\frac{2}{3}}\,.\label{tildeL}
 \ee

\section{Microscopic model}

We now turn to a microscopic description of bridging in sinusoidal slits using the framework of classical Density Functional Theory, where the
central quantity is the one-body fluid density (density profile), $\rho(\rr)$. Among all admissible density profiles, the one corresponding to the
equilibrium state, $\rho_{\rm eq}(\rr)$, is obtained by minimizing a grand potential functional $\Omega[\rho]$ \cite{ebner,evans79}
 \bb
 \Omega[\rho]={\cal F}[\rho]+\int\dd\rr\rhor[V(\rr)-\mu]\,.\label{om}
 \ee
Here, $V(\rr)$ is the external potential, which in our case accounts for the presence of the confining sinusoidal walls, chosen to be purely
repulsive:
 \bb
 V(x,z)=\left\{\begin{array}{cc}
 0\,;&|z|<z_w(x)\,,\\
\infty\,;&{\rm otherwise}\,,
\end{array}\right.\label{ua}
 \ee
with $z_w(x)$ given by (\ref{sin}) and ${\cal F}[\rho]$ is the intrinsic Helmholtz free-energy functional which contains all the information about
the fluid-fluid intermolecular interaction. The formulation of DFT is exact in principle and higher functional derivatives of the free-energy
functional determine a hierarchy of correlation functions. However, the exact form of ${\cal F}[\rho]$ is generally not known, necessitating
approximations.

Although generic models have been developed for a wide range of systems, it is often more practical to tailor the approximation of the intrinsic
free-energy functional to the specific microscopic model defined by the fluid-fluid potential. For simple fluids, reliable perturbative approaches
have been developed in the spirit of the early van der Waals theory. This method separates the fluid-fluid pairwise potential, u(r)u(r) (where rr is
the distance between the centers of two interacting particles), into a purely repulsive part, urep(r)urep?(r), which primarily determines the fluid
structure, and an attractive part, uatt(r)uatt?(r), treated as a perturbation.

Although generic models applicable for a wide range of systems have been developed, it is more common  to tailor the approximation of the intrinsic
free energy functional to the specific microscopic model defined by the fluid-fluid potential. For simple fluids, reliable perturbative approaches
have been developed in the spirit of the early van der Waals theory. This technique is based on a separation of the fluid-fluid pairwise potential,
$u(r)$ (where $r$ is the distance between the centers of the two interacting particles), into a purely repulsive part, $u_{\rm rep}(r)$, which
primarily determines the fluid structure, and an attractive part, $u_{\rm att}(r)$, treated as a perturbation.

To model the interaction between the fluid atoms, we use the truncated, non-shifted Lennard-Jones potential, with the attractive part given by:
 \bb
 u_{\rm att}(r)=\left\{\begin{array}{cc}
 0\,;&r<\sigma\,,\\
4\varepsilon\left[\left(\frac{\sigma}{r}\right)^{12}-\left(\frac{\sigma}{r}\right)^6\right]\,;& \sigma<r<r_{\rm c}\,,\\
0\,;&r>r_{\rm c}\,.
\end{array}\right.\label{ua}
 \ee
with a cut-off distance $r_{\rm c}=2.5\,\sigma$. The harshly repulsive part of the Lennard-Jones potential is mapped to a fluid of hard spheres with
diameter $\sigma$, serving as natural reference system for both homogeneous and inhomogeneous systems \cite{hansen}.

Returning to the perturbative scheme for the intrinsic free energy functional, we separate it  into the ideal gas contribution and an excess part:
 \bb
 {\cal F}[\rho]={\cal F}_{\rm id}[\rho]+{\cal F}_{\rm ex}[\rho]\,,
 \ee
where the ideal part is known exactly:
  \bb
  {\cal F}_{\rm id}[\rho]=\beta^{-1}\int\dr\rho(\rr)\left[\ln(\rhor\Lambda^3)-1\right]\,,
  \ee
with $\Lambda$ being the thermal de Broglie wavelength and $\beta=1/k_{\rm B}T$ the inverse temperature. The excess part of  the intrinsic
free-energy functional is further split into a hard-sphere contribution and an attractive part treated in a  mean-field fashion:
  \bb
  {\cal F}_{\rm ex}[\rho]={\cal F}_{\rm hs}[\rho]+\frac{1}{2}\int\int\dd\rr\dd\rr'\rhor\rho(\rr')u_{\rm att}(|\rr-\rr'|)\,. \label{f}
  \ee

The hard-sphere part of the excess free energy is approximated using Rosenfeld's Fundamental Measure Theory (FMT)  \cite{ros},
 \bb
{\cal F}_{\rm hs}[\rho]=\frac{1}{\beta}\int\dd\rr\,\Phi(\{n_\alpha\})\,,\label{fmt}
 \ee
which accurately captures short-range correlations  between fluid particles and satisfies exact statistical-mechanical sum rules. The free energy
density $\Phi$ depends on the set of weighted densities $\{n_\alpha\}$, which within the original Rosenfeld approach consist of four scalar and two
vector functions given by convolutions of the density profile and the corresponding weight function:
 \bb
 n_\alpha(\rr)=\int\dr'\rho(\rr')w_\alpha(\rr-\rr')\;\;\alpha=\{0,1,2,3,v1,v2\}\,,  \label{weightden}
 \ee
where $w_3(\rr)=\Theta({\cal{R}}-|\rr|)$, $w_2(\rr)=\delta({\cal{R}}-|\rr|)$, $w_1(\rr)=w_2(\rr)/4\pi {\cal{R}}$, $w_0(\rr)=w_2(\rr)/4\pi
{\cal{R}}^2$, $w_{v2}(\rr)=\rr\delta({\cal{R}}-|\rr|)/{|\rr|}$, and $w_{v1}(\rr)=w_{v2}(\rr)/4\pi {\cal{R}}$. Here, $\Theta$ is the Heaviside
function, $\delta$ is the Dirac function, and ${\cal{R}}=\sigma/2$.

\begin{figure*}[htb]
  \includegraphics[width=\linewidth]{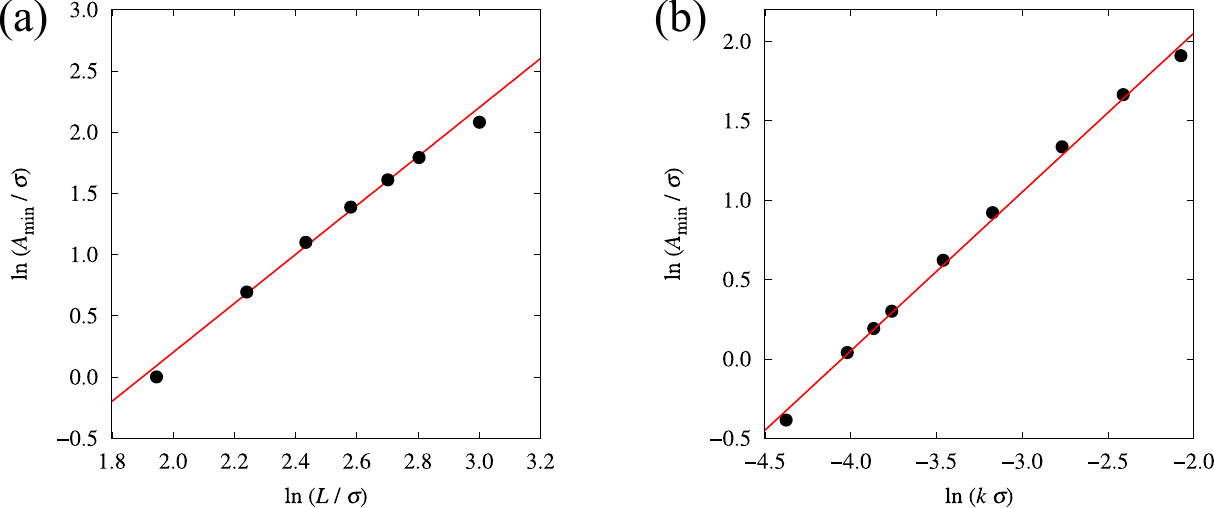}
\caption{DFT results illustrating the scaling properties of the minimal amplitude $A_{\rm min}$ necessary for bridge formation in sinusoidal slits.
a) A log-log plot of $A_{\rm min}$ as a function of mean slit width  $L$. The solid line with a slope of $2$ verifies the expected quadratic scaling,
$A_{\rm min}\propto L^2$, in accordance with Eq.~(\ref{A_min}). The wavelength of the sinusoidal walls is fixed at $\lambda=50\,\sigma$
($k\approx0.13\,\sigma^{-1}$). b)A log-log plot of $A_{\rm min}$ versus wavenumber $k$, with $L=20\,\sigma$ fixed. The solid line with a slope of 1
confirms the linear scaling $A_{\rm min}\propto k$, consistent with Eq.~(\ref{A_min}). All results are obtained at a reduced temperature
$T/T_c=0.92$. } \label{fig_Amin}
\end{figure*}

\begin{figure*}[htb]
\includegraphics[width=\linewidth]{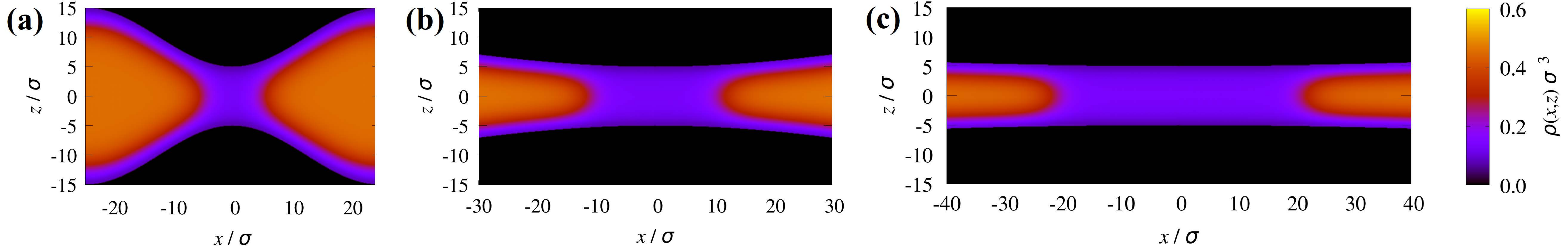}
\caption{Illustrative two-dimensional DFT density profiles, $\rho_{\rm eq}(x,z)$, corresponding to bridged states at the points of bridging
transitions in  sinusoidal slits formed by hard walls with wavelengths of a) $\lambda=50\,\sigma$ ($k\approx0.126\,\sigma^{-1}$), b)
$\lambda=200\,\sigma$ ($k=0.031\,\sigma^{-1}$), and c) $\lambda=500\,\sigma$ ($k=0.013\,\sigma^{-1}$). In all cases, the wall amplitude is
$A=5\,\sigma$, the mean slit separation  is $L=20\,\sigma$, and the temperature is $T/T_c=0.92$.  The plots highlight regions containing a single
gas-like bridge.} \label{fig_dp}
\end{figure*}

\begin{figure}[htb]
\includegraphics[width=\columnwidth]{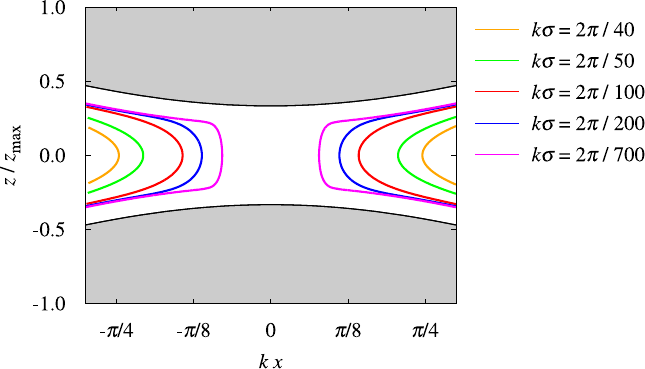}
\caption{Contour plot illustrating the shapes of bridges in sinusoidal slits with varying wavelengths, as obtained from DFT at their respective
bridging transitions. The bridge contours are shown in rescaled dimensionless variables, with $z_{\rm max}=L/2+A=15\,\sigma$ ($L=20\,\sigma$ and
$A=5\,\sigma$). The shrinking of the bridge contours with increasing wavelength suggests that the critical exponent governing the growth of the
bridging films is less than unity. The temperature is fixed at $T/T_c=0.92$. } \label{fig_cont}
\end{figure}

\begin{figure}[htb]
\includegraphics[width=\columnwidth]{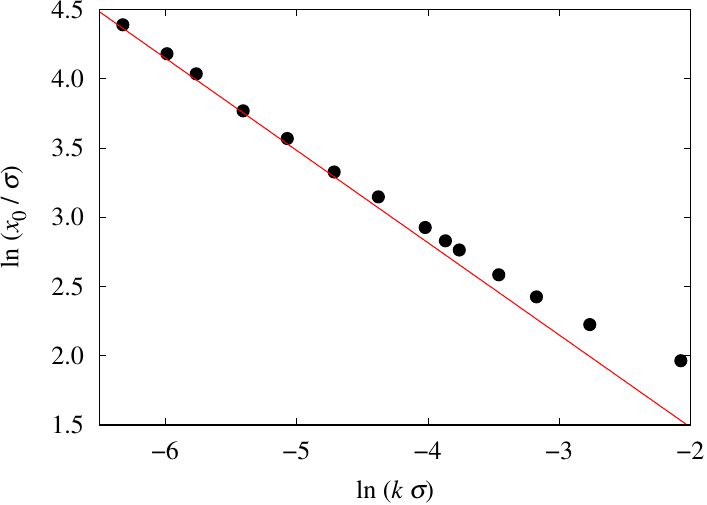}
\caption{Log-log plot displaying numerical DFT results for the growth of the bridging film, characterized by half of its horizontal extension $x_0$,
in sinusoidal slits at the bridging transition as the wave number $k$ decreases.  The wall amplitude and mean separation are fixed at $A=5\,\sigma$
and $L=20\,\sigma$. The straight line with a slope of $-2/3$ confirms the asymptotic prediction described by  Eq.~(\ref{asym_x0}). The temperature is
set to $T/T_c=0.92$.} \label{fig_x0}
\end{figure}

\begin{figure}[htb]
\includegraphics[width=\columnwidth]{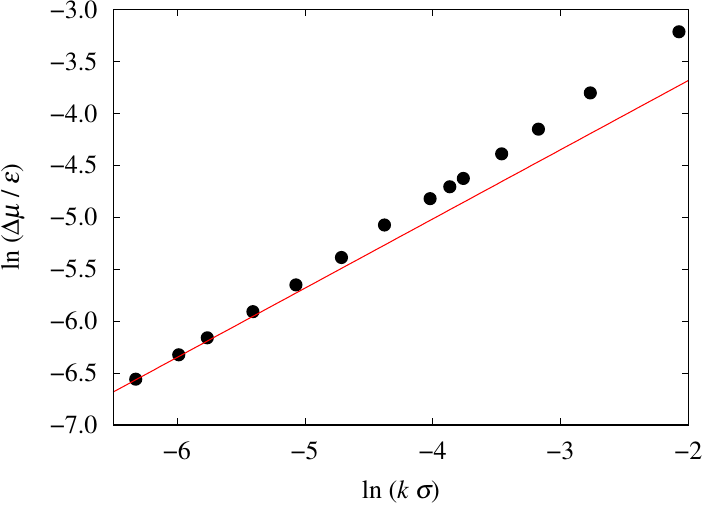}
\caption{Log-log plot illustrating numerical DFT results for the decay of the chemical potential difference $\Delta\mu$ between the bridging
transition in sinusoidal slits and capillary condensation in an infinite planar slit of width $L-2A$ as the wavenumber $k$ decreases. The solid line
with a slope of $2/3$ confirms the expected scaling behavior predicted by Eq.~(\ref{asym_dmu}). The temperature is set to $T/T_c=0.92$.}
\label{fig_dmu}
\end{figure}

Minimization of (\ref{om}) leads to the Euler-Lagrange equation
 \bb
 V(\rr)+\frac{\delta{\cal F}_{\rm hs}}{\delta\rho_{\rm eq}(\rr)}+\int\dd\rr'\rho(\rr')u_{\rm att}(|\rr-\rr'|)=\mu\,,\label{el}
 \ee
which can be recast into a self-consistent equation for the equilibrium density profile:
 \bb
  \rho_{\rm eq}(\rr) = \Lambda^{-3} \exp\left[\beta\mu-\beta V(\rr) + c^{(1)}(\rr)\right]\,. \label{selfconsistent}
 \ee
Here, $c^{(1)}(\rr)=c^{(1)}_\mathrm{hs}(\rr)+c^{(1)}_\mathrm{att}(\rr)$ is the one-body direct correlation function, where the hard-sphere
contribution is
  \bb
  c^{(1)}_{\rm hs}(\rr)= -\sum_\alpha \int\dd\rr'\; \frac{\partial\Phi(\{n_\alpha\})}{\partial n_\alpha} \, w_\alpha(\rr'-\rr)   \label{chs}
 \ee
 and the attractive contribution is
  \bb
  c^{(1)}_\mathrm{att}(\rr) = -\beta\int \dd\rr'\; u_{\rm att}(|\rr-\rr'|)\,\rho_{\rm eq}(\rr')\,.  \label{catt}
 \ee

Eq.~(\ref{selfconsistent}) was solved numerically using Picard's iteration on a two-dimensional rectangular grid with a spacing of $0.01\,\sigma$. In
this way, we determined  a density profile corresponding to a local minimum of the grand potential,  representing either a gas-like or a bridged
configuration. At the point of the bridging transition, $\delta \mu_b$, the grand potentials of the two configurations are equal. The convolutions in
Eqs.~(\ref{weightden}), (\ref{chs}), and (\ref{catt}) were evaluated using the Fourier transform (see \cite{our_sin} for details).

\section{Results}

In this section, we present the DFT results obtained from the model described earlier, aiming to test the scaling and asymptotic properties of the
bridging transition in sinusoidal slits, as predicted in Section \ref{macro}. The external potential $V(\rr)$ is chosen to model sinusoidally-shaped
hard walls, corresponding to a Young's contact angle $\theta=\pi$; this is equivalent to a complete wetting scenario ($\theta=0$) with the roles of
gas and liquid phases reversed. Thus, we  consider supersaturated systems $\mu>\mu_{\rm sat}$ (i.e., $\delta\mu<0$), where the bulk reservoir is in a
liquid state, and the bridges are gas-like.

All the presented results are expressed in terms of the microscopic parameters $\sigma$ and $\varepsilon$, which set as the length and energy units,
respectively. The DFT calculations correspond to a temperature $T=0.92\,T_c$, for which the chemical potential at saturation is $\mu_{\rm
sat}=-3.9435\,\varepsilon$. The corresponding one-body bulk densities of the coexisting  gas and liquid phases are $\rho_g=0.1040\,\sigma^{-3}$ and
$\rho_l=0.4313\,\sigma^{-3}$, with a surface tension between them of $\gamma= 0.0562\,\varepsilon\sigma^{-2}$. The critical temperature of our fluid
model is $k_bT_c\approx1.41\,\varepsilon$.

We first test the scaling properties of the minimal amplitude $A_{\rm min}$ required for the bridging transition. Fig.~\ref{fig_Amin}a shows the DFT
results for $A_{\rm min}$ across various pore widths,  with the wavelength of the walls fixed at $\lambda=50\,\sigma$ (hence
$k\approx0.13\,\sigma^{-1}$). The results, presented in a log-log plot, confirm the expected dependence of $A_{\rm min}\propto L^2$ as predicted by
Eq.~(\ref{A_min}). Similarly, the expected linear scaling of $A_{\rm min}$ with $k$ is verified in Fig.~\ref{fig_Amin}b, where the mean pore width
was fixed at $L=20\,\sigma$.

Next, we examine the behaviour of the bridging transition associated with increasing the system's wavelength to test the power-law dependence
(\ref{asym_x0}) in the limit of $k\to0$. In Fig.~\ref{fig_dp}, we present illustrative two-dimensional density profiles in the $x$-$z$ projection of
bridges formed at the transition for systems with different wavelengths, while keeping the mean pore width and the amplitude of the walls fixed
$L=20\,\sigma$ and $A=5\,\sigma$, respectively. We observe that  gas-like bridges form around the necks, with their width increasing as $\lambda$
increases (i.e., as $k$ decreases).

In Fig.~\ref{fig_cont}, we compare the contours of bridges for additional systems with varying wavelengths, using rescaled variables $x'=x/A$ and
$z'=kz$.  This rescaling allows for a direct comparison within a single plot; here, the contours are defined as the set of points $(x_c,z_c)$
corresponding to the midpoints of the liquid and gas bulk densities, $\rho(x_c,z_c)=(\rho_g+\rho_l)/2$. We observe that, in these rescaled
coordinates, the bridges shrink with increasing period, indicating that the critical exponent associated with the divergence of the bridging width is
less than unity.

To determine this exponent, a log-log plot is presented in Fig.~\ref{fig_x0}, where the dependence of $x_0$ on $k$ for systems with the periods up to
$\lambda=3500\,\sigma$ is shown. For a comparison, a line with the slope of $-2/3$ is also displayed, demonstrating  that the asymptotic prediction
given by Eq.~(\ref{asym_x0}) is followed almost perfectly.

Finally, we test the expected relation between  the bridging transition in sinusoidal slits and capillary condensation in infinite planar slits.
According to Eq.~(\ref{asym_dmu}), the locations of the two transitions converge as $k$ decreases, such that the difference in the chemical
potentials corresponding to the respective transitions, $\Delta\mu\equiv\delta\mu_b-\delta\mu_{cc}(L-2A)$, vanishes in the limit  $k\to0$ as
$k^{2/3}$.  This behaviour is indeed observed within the DFT model, as shown in Fig.~\ref{fig_dmu}, confirming that  $\Delta\mu$  asymptotically
approaches a slope of $2/3$.

\section{Concluding remarks}

In this work, we have investigated the asymptotic properties of bridging transitions that occur between two sinusoidally shaped walls  brought close
together in a reflection-symmetric configuration. The system is characterized by three geometric parameters: the amplitude, $A$, the wavenumber, $k$
(or equivalently  the wavelength $\lambda$), and the mean separation between the walls, $L$. We began by formulating a Kelvin-like equation
appropriate for this model, which determines the chemical potential (or pressure) at which the bridging transition occurs. To achieve this, we
utilized the concept of the edge contact angle, $\theta_e$,  mapping the condensed part of the system to a planar slit of finite length, where
$\theta_e$ serves as the angle subtended between the menisci formed in the condensed state and the walls. This mapping allows us to relate this angle
to the geometry of the walls.

Subsequently, we focused on the case of weakly corrugated walls by linearizing the Kelvin equation (\ref{kelvin_sin_full})  with respect to the
roughness parameter $Ak$, leading to the analytically tractable equation (\ref{kelvin_sin}).  The analysis of the bridging transition based on on
this linearized equation can be summarized as follows:

\begin{enumerate}

\item {\bf Critical Phenomenon:} There exists a critical behavior associated with  a continuous divergence in the extension of the bridges
as the walls flatten, such that $k\to0$  (i.e., the wavelength tends to infinity). The critical exponent describing this divergence is $2/3$,
indicating a universal yet geometry-specific scaling. This implies that bridges formed at the necks of sinusoidal slits shrink relative to the
system's wavelength.

\item {\bf  Asymptotic Approach to Capillary Condensation:} As the system's wavelength increases, the location of the bridging transition approaches
that of capillary condensation in an infinite planar slit of width $L-2A$, with the difference in the chemical potential scaling as $\sim k^{2/3}$.

\item {\bf Covariance Relation:} This behaviour suggests a covariance between bridging transitions and capillary condensation,
which can be expressed through an appropriate relation involving the respective geometric parameters.

\item {\bf  Amplitude-Dependent Transition:} Reducing the system's amplitude $A$ (while keeping $k$ fixed) results in a qualitatively different scenario.
In this case, the system hits a marginal amplitude $A_{\rm min}$, below which bridge formation is not possible. $A_{\rm min}$ scales linearly with
$k$ and quadratically with $L$.

\item {\bf  Stability of Bridging Transition:} For $A>A_{\rm min}$, the bridging transition is always stable and occurs prior to capillary condensation.

\end{enumerate}

The last point may initially seem counterintuitive. One might expect that, in some cases, the fusion of the bridges to form a single liquid-like
phase throughout the entire system would be energetically favorable. Specifically, this scenario might be anticipated in two cases:  (i) vanishing
roughness,  and ii) large $L$, where the surface free-energy cost due to menisci is substantial. However, there is no contradiction. For the former
case  ($k\to0$), Fig.~\ref{fig_cont} shows that the relative volume of the bridges diminishes with narrowing  walls, implying that capillary
condensation requires a progressively larger amount of the metastable phase. The corresponding volume free-energy cost (per period) scales as
$L\lambda$, thus increases as $\sim k^{-1}$, a rate that exceeds the growth of the volume term required for bridging. For large values of $L$, note
that the minimal amplitude $A_{\rm min}$ permitting bridging transition must also be large (cf. the point 4 above), making it unsurprising that
bridging remains favorable.

It should be emphasized, however, that our conclusions apply strictly within the regime of weakly corrugated walls (as described by the linearized
Kelvin equation). Increasing $L$ necessitates both an increase in $A$ (to satisfy the bridging condition) and a decrease in $k$ (to remain in the
weakly corrugated regime). Nonetheless,  the condition $A>A_{\rm min}$ cannot always be satisfied since $A$ is naturally limited by $L/2$. Thus, for
a given system wavelength, there exists a maximum mean separation $L_{\rm max}$ -- scaling linearly with $\lambda$ -- beyond which  $A>A_{\rm min}$
is never fulfilled, and the system only experiences capillary condensation.

In summary, the objective of this work was to explore the properties of bridging transitions in sinusoidal slits as the roughness of the walls is
reduced. We have shown that the limits  $A\to0$ and $k\to0$  must be distinguished, as only in the latter case does the bridging transition exhibit a
critical phenomenon characterized by the divergence of the bridging film width.  In contrast, the limit $A\to0$ cannot be considered, since  a
minimal amplitude $A_{\rm min}$ is required, below which the system exhibits only capillary condensation. The scaling properties of the asymptotic
behaviour of bridging and  $A_{\rm min}$ have been determined and supported by microscopic DFT results.

It should be noted that, strictly speaking, the bridging transition is generally rounded due to interfacial fluctuations. However, this rounding
becomes negligible as the bridging films grow and becomes irrelevant in the $k\to0$ limit. In this study, we restricted ourselves to completely wet
walls ($\theta=0$), but extending the analysis to partial wetting ($\theta>0$) is straightforward. In fact, the Kelvin equation is expected to be
more accurate in this case due to the absence of wetting/drying films adsorbed on the walls, which were not considered in our model.

Potential extensions of the present work include exploring bridging phenomena beyond the small-corrugation regime, where interactions with other
surface phenomena are anticipated. This includes the \emph{osculation transition} \cite{osc, osc2}, which refers to the macroscopic filling of wall
troughs that can be either continuous or first-order depending on temperature. Additionally, it would be interesting to compare the current results
with those obtained for axially symmetric models (such as sinusoidally undulated cylindrical pores) and to investigate commensurate effects and their
impact on phase behavior when the fluid particles are non-spherical.

\begin{acknowledgments}
\noindent This work was financially supported by the Czech Science Foundation, Project No. 21-27338S.
\end{acknowledgments}


\begin{thebibliography}{99}

\bibitem{evans90}
R. Evans, J. Phys.: Condens. Matter {\bf 2}, 8989 (1990).

\bibitem{hend}
D. Henderson, {\it Fundamentals of Inhomoheneous Fluids}, Marcel Dekker, New York (1992).

\bibitem{NF81}
M. E. Fisher and H. Nakanishi, J. Chem. Phys. {\bf 75}, 5857 (1981).

\bibitem{NF83}
H. Nakanishi and M. E. Fisher, J. Chem. Phys. {\bf 78}, 3279 (1983).

\bibitem{derj}
B. V. Derjaguin, Acta Phys. Chem. {\bf 12}, 181 (1940).

\bibitem{sul79}
D. E. Sullivan, Phys. Rev. B {\bf 20},3991 (1979).

\bibitem{sul81}
D. E. Sullivan, J. Chem. Phys. {\bf 74}, 2604 (1981).

\bibitem{evans86}
R. Evans, U. Marini Bettolo Marconi, and P. Tarazona, J. Chem. Phys.  {\bf 84}, 2376 (1986).

\bibitem{binder08}
K. Binder, J. Horbach, R. Vink, and A. De Virgiliis, Soft Matter {\bf 4}, 1555 (2008).

\bibitem{gelb}
L. D. Gelb, K. E. Gubbins, R. Radhakrishnan, and M. Sliwinska-Bartkowiak, Rep. Prog. Phys. {\bf 62}, 1573 (1999).

\bibitem{geim}
Q. Yang, P. Z. Sun, L. Fumagalli, Y. V. Stebunov, S. J. Haigh, Z.W. Zhou, I. V. Grigorieva, F. C. Wang, and A. K. Geim, Nature (London) {\bf 588},
250 (2020).

\bibitem{privman}
V. Privman and M. E. Fisher, J. Stat. Phys. {\bf 33}, 385 (1983).

\bibitem{faucher}
S. Faucher et al., J. Phys. Chem. C {\bf 123}, 21309 (2019).

\bibitem{aluru}
N. R. Aluru et al., Chem. Rev. {\bf 123}, 2737 (2023).

\bibitem{araujo}
N. A. M. Ara\'ujo, Soft Matter {\bf 19}, 1695 (2023).


%bridging

\bibitem{israel}
 J. N Israelachvili, Intermolecular and Surface Forces, 2nd ed.; Academic
Press: London, (1992).

\bibitem{mason}
G. Mason and W. C. Clark, Chem. Eng. Sci. {\bf 20}, 859 (1965).

\bibitem{beysens}
 D. Beysens, D. and D Esteve, Phys. Rev. Lett. {\bf 54}, 2123 (1985).

\bibitem{willett}
C. D. Willett, M. J. Adams, S. A. Johnson, and J. P. K. Seville,  Langmuir {\bf 16}, 9396 (2000).

\bibitem{kohonen}
M. Kohonen, D. Geromichalos, M. Scheel, C. Schier, and S. Herminghaus,  Phys. A {\bf 7}, 339 (2004).

\bibitem{fournier}
Z. Fournier at. al, J. Phys.: Condens. Matter {\bf 17}, S477 (2005).


\bibitem{gogelein}
C. Gogelein, M. Brinkmann, M. Schroter, and S. Herminghaus, Langmuir {\bf 26}, 17184 (2010).



\bibitem{macro_rings}
F. M. Orr, L. E. Scriven, and A. P. Rivas,  J . Fluid Mech. {\bf 67}, 723 (1975).

\bibitem{macro_vogel}
T. I. Vogel, Pacific J. Math. {\bf 224}, 367 (2006).

\bibitem{macro_willet}
C. D. Willett, M. J. Adams, S. A. Johnson, and J. P. K. Seville, Langmuir {\bf 16}, 9396 (2000).

\bibitem{yeomans}
H. T. Dobbs, G. A. Darbellay, and J. M. Yeomans, Europhys. Lett. {\bf 18}, 439 (1992).

\bibitem{yeomans2}
H. T. Dobbs and J. M. Yeomans, J. Phys.: Condens. Matter {\bf 4}, 10133 (1992).


\bibitem{bauer}
C. Bauer, T. Bieker, and S. Dietrich, Phys. Rev. E {\bf 62}, 5324 (2000).

\bibitem{andrienko}
D. Andrienko, P. Patr\'\i cio, and O. I. Vinogradova, J. Chem. Phys. {\bf 121}, 4414 (2004).

\bibitem{grof}
Z. Grof and C. J. Lawrence, and F. \v St\v ep\'anek, J. Colloid Interface Sci. {\bf 319}, 182 (2008).

\bibitem{hopkins}
 P. Hopkins, A. J. Archer, and R. Evans, J. Chem. Phys, {\bf 131}, 124704 (2009).

\bibitem{dutka}
F. Dutka and M. Napi\'orkowski, J. Chem. Phys. {\bf 133}, 051101 (2010).

\bibitem{cheng}
T.-L. Cheng and Y. U. Wang, Langmuir {\bf 28}, 2696 (2012).

\bibitem{mal}
A. Malijevsk\'y, Mol. Phys. {\bf 113}, 1170 (2015).

\bibitem{mal_par}
 A. Malijevsk\'y and A. O. Parry, Phys. Rev. E {\bf 92}, 022407 (2015).

\bibitem{farmer}
T. P. Farmer and J. C. Bird, J. Colloid Interface Sci. 454, 192 (2015).

\bibitem{labbe}
M. Labb\'e-Laurent, A. D. Law, and S. Dietrich, J. Chem. Phys. {\bf 147}, 104701 (2017).

\bibitem{vasil}
O. A. Vasilyev, S. Dietrich, and S. Kondrat, Soft Matter {\bf 14}, 586 (2018).

\bibitem{vasilyev}
O. A. Vasilyev, M. Labb\'e-Laurent, S. Dietrich, and S. Kondrat, J. Chem. Phys. {\bf 153}, 014901 (2020).

\bibitem{mal24}
A. Malijevsk\'y and M. Posp\'\i\v sil, Phys. Rev. E {\bf 109}, 034801 (2024).

\bibitem{our_slit}
A. Malijevsk\'y, A. O. Parry, and M. Posp\'\i\v sil, Phys. Rev. E {\bf 96}, 020801(R) (2017).

\bibitem{open_slits}
A. Malijevsk\'y and A. O. Parry,  Phys. Rev. E {\bf 104}, 044801 (2021).


\bibitem{ebner}
C. Ebner and W. F. Saam, Phys. Rev. Lett. {\bf 38}, 1486 (1977).

\bibitem{evans79}
R. Evans, Adv. Phys. {\bf 28}, 143 (1979).

\bibitem{hansen}
J. P. Hansen and J. R. McDonald, Theory of Simple Liquids (Academic, New York, 2005) 3rd ed.

\bibitem{ros}
Y. Rosenfeld,  Phys. Rev. Lett. {\bf 63}, 980 (1989).

\bibitem{our_sin}
M. Posp\'\i\v sil and A. Malijevsk\'y, Phys. Rev. E {\bf 106}, 024801 (2022).

\bibitem{osc}
M. Posp\'\i\v sil, A. O. Parry  and A. Malijevsk\'y, Phys. Rev. E {\bf 105}, 064801 (2022).

\bibitem{osc2}
 A. O. Parry , M. Posp\'\i\v sil, and A. Malijevsk\'y, Phys. Rev. E {\bf 106}, 054802 (2022).











\end{thebibliography}
\end{document}